\input{psfig}
\input{epsf}

\def\mev{\,{\rm Me\kern-0.1em V}}
\def\gev{\,{\rm Ge\kern-0.1em V}}

\documentstyle[12pt]{article}

\begin{document}
\vspace*{-1.25in}
\small{
\begin{flushright}
FERMILAB-PUB-98/159-T \\[-.1in] 
May, 1998 \\
\end{flushright}}
\vspace*{.75in}
\begin{center}
{\Large{\bf  Quenched Chiral Artifacts for Wilson-Dirac Fermions}}\\
\vspace*{.45in}
{\large{W. ~Bardeen$^1$,
A.~Duncan$^2$, 
E.~Eichten$^1$,  and
H.~Thacker$^3$}} \\ 
\vspace*{.15in}
$^1$Fermilab, P.O. Box 500, Batavia, IL 60510 \\
$^2$Dept. of Physics and Astronomy, Univ. of Pittsburgh, 
Pittsburgh, PA 15260\\
$^3$Dept.of Physics, University of Virginia, Charlottesville, 
VA 22901
\end{center}
\vspace*{.3in}
\begin{abstract}
We examine artifacts associated with the chiral symmetry
 breaking induced through the use of Wilson-Dirac
 fermions in lattice Monte Carlo computations.   For
 light quark masses, the conventional quenched theory can not be
 defined using direct Monte Carlo methods due to the
 existence of nonintegrable poles in physical quantities.
 These poles are associated with the real eigenvalue spectrum of
 the Wilson-Dirac operator.   We show how this
 singularity structure can be observed in the analysis of
 both QED in two dimensions and QCD in four
 dimensions.
\end{abstract}

\newpage
\section{Introduction}

	During the past twenty years, considerable progress has
been achieved in studying the nonperturbative structure of 
gauge field theory through numerical computations of lattice 
field theory.   The principal method of analysis has involved 
Monte Carlo computations of physical quantities such as 
meson correlation functions where fermions have been 
treated using the quenched approximation.   The quenched 
approximation omits the contribution of the fermion 
determinant to the functional integral defining the transition 
amplitudes.  This approximation is known to be quite 
sensitive to the particular formulation used to define 
fermions on the lattice.   In this paper we will address certain 
problems associated with the study of light fermions using 
the Wilson-Dirac formulation \cite{Wilson} of lattice fermions.

	In the continuum the usual Dirac operator defining the 
fermion action preserves a chiral symmetry structure in the 
presence of gauge interactions which is broken only by the 
addition of fermion mass terms or Yukawa interactions.   In 
the Wilson-Dirac formulation of lattice fermions, this chiral 
structure is explicitly broken by the analogue of second derivative 
terms which are needed to remove the doubling degeneracy 
of the naive lattice action for Dirac fermions.   This chiral 
symmetry breaking modifies the eigenvalue spectrum of the 
Wilson-Dirac operator from that expected for the continuum 
Dirac operator.   In Euclidean space, the chiral structure of the 
usual Dirac operator implies a purely imaginary spectrum for 
its eigenvalues.   However, the Wilson-Dirac operator will 
have a spectrum of complex eigenvalues which fill a region 
of the complex plane \cite{SmitVink,Vink}.   In 
particular, there will be a spread in the values of the real part 
of these eigenvalues which means that the massless fermion 
limit cannot be uniquely specified and the chiral limit can 
only be defined through ensemble averages of physical 
quantities.   Unfortunately, the relevant physical quantities, 
such as the pion correlator, are singular functions of the 
Wilson-Dirac eigenvalues in the standard quenched 
approximation and there is growing evidence \cite{QED2}
that ensemble averages may not exist for 
sufficiently small fermion masses.   We will give explicit 
evidence that this is the case for Wilson fermion theories in two and 
four dimensions including QED2 and QCD4.

     In previous work \cite{QED2,MQA} 
we have shown that the precisely real eigenvalues of 
the Wilson-Dirac operator play a special role in the quenched 
theory.   These eigenvalues are the analogue of the zero 
modes of the continuum theory which are associated with the
topological structure of the background gauge field.    In the 
lattice formulation of the theory, these modes are responsible 
for the singular behavior of the quenched theory for light 
fermion masses.    The singularities arise when the fermion 
mass is chosen to lie within the band of real eigenvalues 
which exists because of the chiral symmetry breaking property 
of the Wilson-Dirac operator.    This is well known and is the 
principal reason for the  ``exceptional" configurations 
observed in  Monte Carlo calculations of the quenched 
theory \cite{exceptional}.   Indeed, we have previously made a careful 
analysis \cite{QED2} of the correlation between the behavior of the real 
eigenvalue spectrum and the observation of exceptional 
configurations.  However, the problem associated with these 
exceptional configurations goes beyond the annoying 
problem of an occasional large contribution of a single 
configuration and its implication for the statistical errors of 
the associated ensemble averages.     We have previously 
pointed out that this problem cannot be solved  simply by
accumulating larger statistical samples as the exceptional 
configurations are expected to occur at a level where, 
even in large statistical samples, the statistical errors will 
not diminish even in the limit of infinite statistics.

	The singularities associated with the real eigenvalues 
are related to both the chiral symmetry breaking associated 
with the Wilson-Dirac operator and the quenched 
approximation which enhances singularities of the fermion 
propagators that are normally suppressed in the full 
unquenched theory.   The fermion propagator can be written 
as a sum over the eigenvalues of the Wilson-Dirac operator, 
$\lambda_{i}$,
\begin{equation}
	S_{F}(x,y,A(x)) = \sum_{i} v_{i}(x,A)w_{i}(y,A)/(\lambda_i + m_0)
\end{equation}
where $v$ (resp. $w$) are left (resp. right)  eigenmodes in a particular 
background gauge field, $A(x)$.  

In the continuum, the fermion 
mass parameter being nonzero implies that there are no 
singular terms in the sum when the gauge fields are 
smoothly varied.   However, this is not the case on the lattice.   
As the background gauge field is varied, isolated real 
eigenvalues are restricted to remain on the real axis.   The 
symmetries of the Wilson-Dirac operator \cite{QED2,Gattringer}  
imply that the complex eigenvalues occur in complex 
conjugate pairs and a single real eigenvalue cannot become 
two complex-conjugate eigenvalues via smooth variation of 
the gauge potentials.    Because of this constraint, the 
contribution of the real eigenvalue modes to the functional 
integral over the gauge potential can be reduced to a one-
dimensional integral corresponding to the position of the real 
eigenvalue.   Hence, the quenched functional integral will 
contain nonintegrable singularities associated with the 
contribution of the real modes if the fermion mass is chosen 
to lie within the band associated with the spread of the real 
eigenvalues.   For example, the nonsinglet meson propagator 
is obtained by squaring the quark propagator and, 
therefore, the real modes will generate a dipole contribution 
to the meson correlator.    If the bare fermion mass is chosen to lie 
within the band of real eigenvalues, the Monte Carlo 
calculation will smoothly sample eigenvalues in the 
neighborhood of fermion mass corresponding to a 
one-dimensional integration parameter even though the original 
gauge integration corresponds to a very high dimensional 
space of integration variables.    Hence, the sum for the 
meson correlator effectively reduces to
\begin{equation}
	\int {\cal D}A \frac{{\rm Res}(A)}{(\lambda_{\rm pole}(A) + m_o)^2} 
	\rightarrow \int d\lambda \frac{{\rm Res}(\lambda)}{(\lambda + m_o)^2}
\end{equation}
which is not directly defined.   For this reason we argue that 
the naive quenched theory does not exist.   The ``exceptional" 
configurations observed in Monte Carlo calculations are not 
strictly speaking an
exceptional feature but are, instead, a {\em generic} property of the 
quenched theory signaling the breakdown of the naive 
theory for light fermion masses. The Monte Carlo 
averages simply do not converge for sufficiently large 
statistical samples.   Indeed, this singular behavior of the 
naive quenched theory led us to propose \cite{MQA} a modification of 
the quenched theory, MQA, where these  lattice fermion 
artifacts are explicitly removed in the evaluation of the 
quenched correlation functions by restoring the chiral 
structure of the continuum theory.   Others have recently 
proposed that the entire band of light fermion masses might  be 
associated with a massless phase of the theory \cite{Florida}.   
Unfortunately, we shall argue below
that this interpretation cannot be directly applied 
because the naive quenched hadronic correlation functions simply are not 
defined in this region for the reasons given above.

  While we have given a general argument concerning 
the difficulties of defining the naive quenched theory, we 
will now show explicit evidence of this singular behavior in 
a number of calculations in QED2 and QCD4. In particular, we 
wish to show in this paper that nonintegrable singularities in the quenched 
functional integral have a characteristic and clearly detectable
statistical signature 
in a Monte Carlo simulation. In Section 2, the singularity structure
of quenched 2-dimensional QED is explored using a variety of detailed
statistical signatures. Explicit numerical evidence is provided to
show that the quenched functional integral of physical correlators
contains nonintegrable singularities and is therefore undefined. A
convenient statistical diagnostic for revealing the existence and strength
of nonintegrable singularities in a quenched path integral is introduced
and applied. Section 3 illustrates certain universal aspects of the 
statistical behavior of the Monte Carlo simulation 
of divergent integrals using a simple one-dimensional
test function with essentially identical behavior to meson correlators in
quenched lattice gauge theory. 
The analysis of Section 2 for QED2 is repeated in Section 4
using data from a full 4 dimensional quenched QCD simulation at relatively
strong coupling ($\beta$=5.7) - one infected with frequent ``exceptional"
configurations. In Section 5 we comment on the use of spectral characteristics
of the hermitian Wilson-Dirac operator in studying new phases of quenched 
gauge theory with Wilson fermions, and emphasize the need for a regularized
version of quenched theory in studying otherwise undefined correlator averages.

\section{Quenched Singularity Structure of QED2}

   The two-dimensional version of massive quantum electrodynamics,
the massive Schwinger model, has frequently been used as a testbed
for studying the structure of fermions in lattice field theory
\cite{SmitVink}.
It shares many features in common with four dimensional quantum
chromodynamics, including aspects of chiral symmetry, a nontrivial topological
structure, fermion doubling problems on the lattice, and the
exceptional configuration artifacts of Wilson-Dirac fermions.

  In a previous paper \cite{QED2}, we have emphasized the role that 
exactly real eigenvalues play in generating exceptional configurations.
In QED2 with Wilson-Dirac fermions, the real eigenvalues occur in
bands rather than at a particular critical value associated 
with the zero modes of the continuum theory. The bands tend to narrow
for gauge configurations generated with larger values of $\beta$
corresponding to the approach to the continuum theory.

\begin{figure}
\psfig{figure=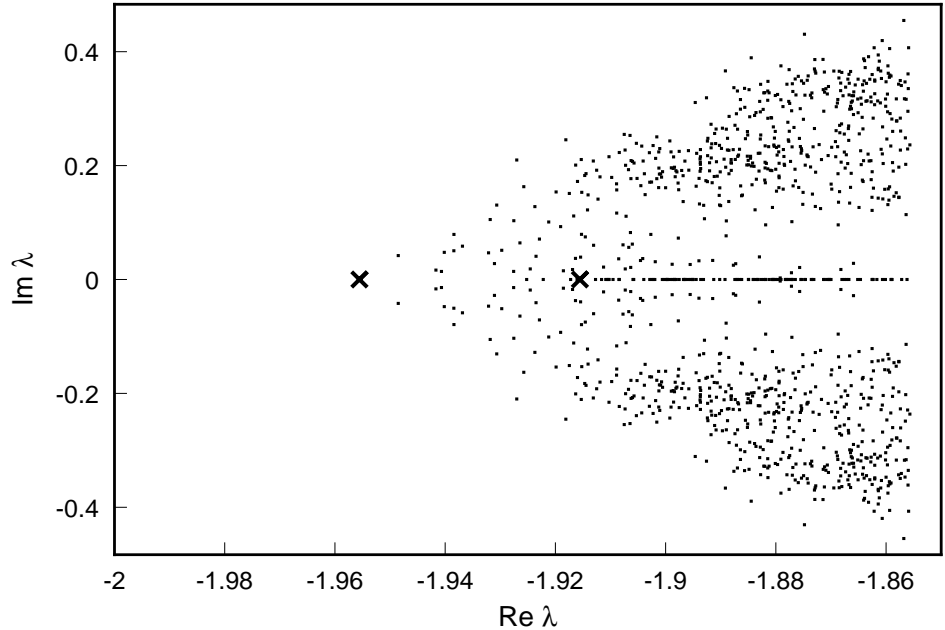,
width=0.95\hsize}
\caption{Wilson-Dirac spectrum near the left 
critical branch [QED2]. The eigenvalue positions corresponding 
to $m_q = 0.10$ and $m_q = 0.06$ are denoted by crosses.}
\label{fig:qed2spec}
\end{figure}
 We have carried out  simulations of QED2 defined on a 10x10
lattice at $\beta$= 4.5. According to the
analysis of Smit and Vink \cite{SmitVink}, the massless theory is
associated with a bare fermion mass given by the approximate relation
$m_{0}=m_{c}\simeq$2.0-0.65/$\beta=$ 1.9155. In Fig(1), we show the spectrum
of accumulated eigenvalues of the Wilson-Dirac operator
in a typical set of 500 consecutive (uncorrelated) configurations in
a quenched simulation at $\beta$=4.5. Only eigenvalues with a real part
less than the critical value of Smit and Vink are shown in this plot.
Both real and complex eigenvalues exist in this region. Because the
real eigenvalues are less than the critical value, they will produce
poles in the fermion propagators at positive values of the fermion
mass, leading to a singular behavior of physical quantities constructed
from these propagators in the quenched theory. We have marked the positions
where real eigenvalues would produce poles for $m_q\equiv m_{0}-m_c$=0.06 and
0.10, respectively. The smaller fermion mass, $m_q$=0.06, lies within the 
band of real eigenvalues, while the larger mass, $m_q$=0.10, lies outside
this region and should not be associated with singular behavior.

   In the quenched theory, various physical quantities, such as correlation
functions, can be computed from products of the valence fermion
propagator. The propagators in turn can be written as  a spectral sum
involving eigenfunctions of the Wilson-Dirac operator and associated 
eigenvalues as in Eq(1). The pseudoscalar (``pion'') 
propagator is then given as a sum over the eigenvalues:
\begin{equation}
 J_{55}(x,y)=\sum_{ij}\frac{{\rm Tr}\gamma_{5}r_{ij}(x,y,A)\gamma_{5}r_{ji}(y,x,A)}{(\lambda_{i}(A)+m_{0})(\lambda_{j}(A)+m_{0})}
\end{equation}
where $m_{0}= m_{q}+m_{c}$. The eigenvalue dependence of the denominators
exhibits the potential for double pole singular behavior in an integral 
over $A$ (for heavy-light mesons, one would encounter a single pole, for
light-light-light baryons in QCD4, a {\em triple} pole in the quenched 
functional integral). 

  As an example of this divergence, we computed the pseudoscalar propagator
at Euclidean time t=2 for $m_q$=0.06 and 0.10. In Fig(2) the cumulative
averages and associated statistical errors are plotted for a sample
of 1000 quenched gauge configurations separated by 10 Metropolis sweeps.
The propagator averages for the lighter mass clearly reflect the presence
of singularities in the functional integral. After {\em apparent} convergence
with the first 100 configurations, 
the contributions of a number of exceptional 
configurations become dominant and no reliable average 
value can be extracted from the simulation. For this mass value, 
we are within the band of real eigenvalues and the 
Monte Carlo process eventually samples
eigenvalues arbitrarily close to the double pole singularity in the
pseudoscalar propagator. We can compare  this behavior to the propagator
average for the larger mass, $m_q$=0.10, which lies outside the band
of real eigenvalues (see Fig(3)). Here the 
propagator appears to converge to its
average value after about 400-500 sweeps and there is 
no obvious singular behavior. 

\begin{figure}
\psfig{figure=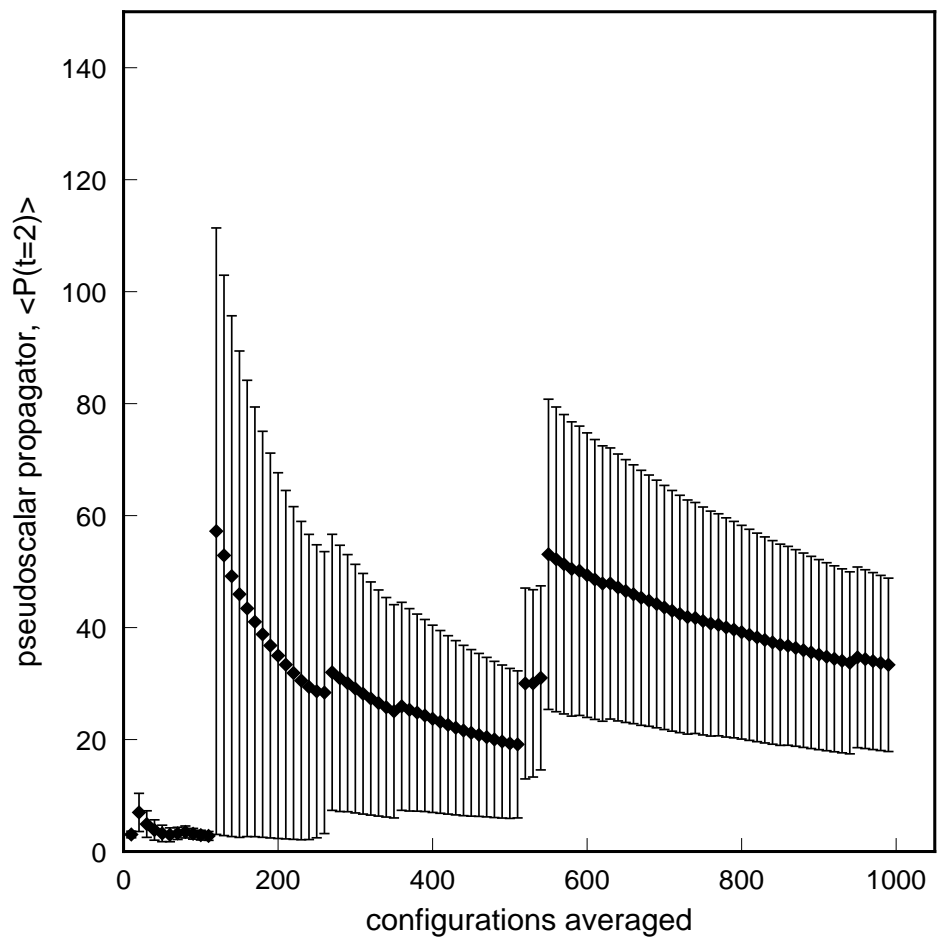,
width=0.95\hsize}
\caption{Convergence of pseudoscalar correlator with increasing statistics
in quenched QED2. Here $m_q$=0.06.}
\label{fig:qedqnchconv0.06}
\end{figure}

\begin{figure}
\psfig{figure=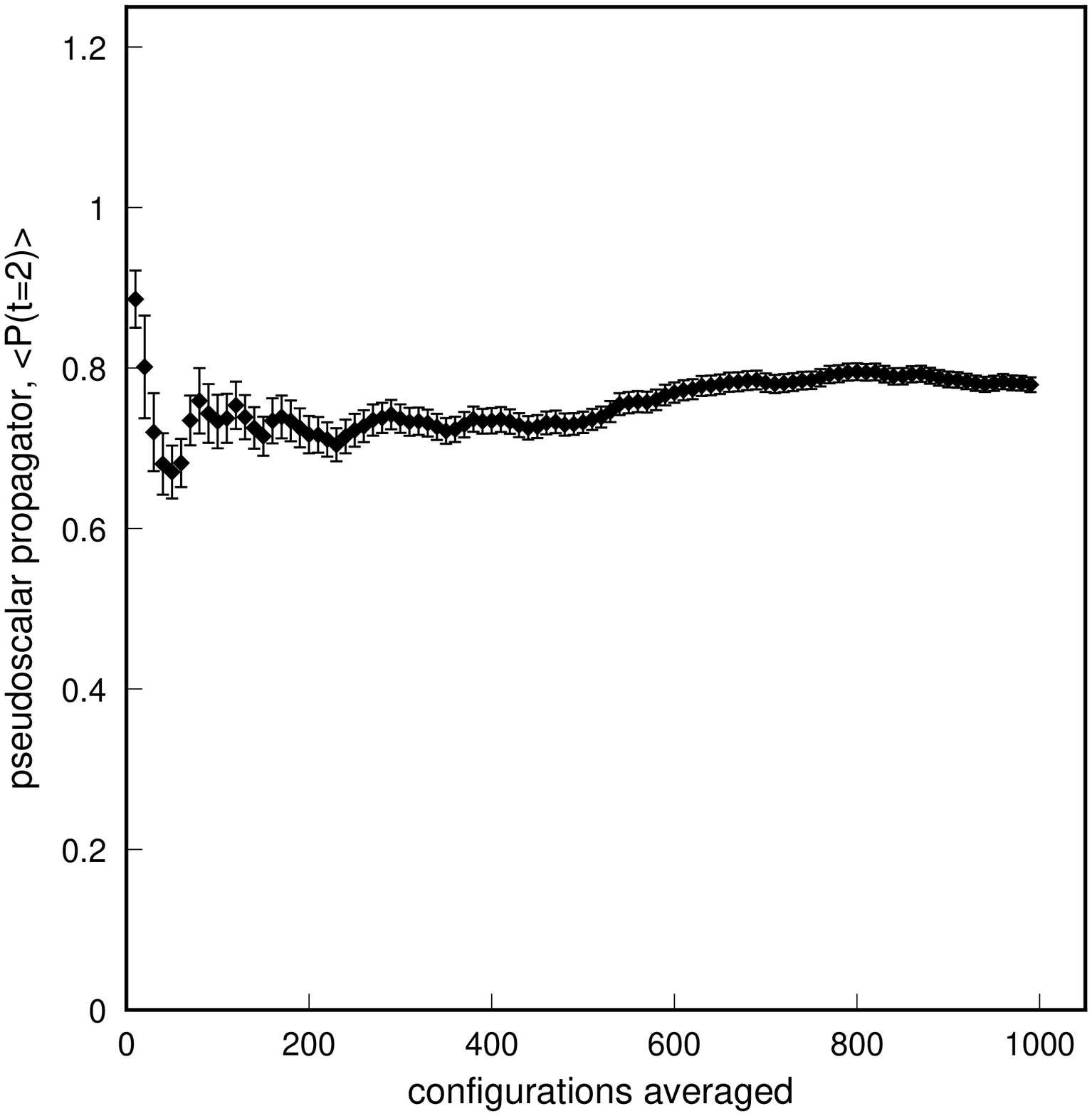,
width=0.95\hsize}
\caption{Convergence of pseudoscalar correlator with increasing statisitics
in quenched QED2. Here $m_q$=0.10.}
\label{fig:qedqnchconv0.10}
\end{figure}

  Even if we do not directly know the eigenvalue spectrum (and this is
 in general the case for QCD in 4 dimensions), we can infer the singular
 structure of the quenched integral directly from physical quantities
 such as meson propagators. For example, the pseudoscalar propagator in
 Eq(3) is a singular function as an eigenvalue approaches the critical
 value $\lambda_{i}\rightarrow m_{0}$. In this singular limit, one
 eigenvalue will dominate the propagator sum in the form of a double pole
 singularity. The Monte Carlo average will sample eigenvalues in the
 neighborhood of this singularity so long as the mass lies within the 
 band defined by the purely real part of the spectrum. It should be expected
 that eigenvalues close to the singularity will be uniformly sampled
 (given sufficient statistics) as neighboring gauge configurations smoothly
 vary the values of the real eigenvalues. 
 (This uniform sampling is a consequence of the fact that, unlike full QCD,
 the distribution of quenched gauge configurations is unaffected by singularities 
 in the quark propagator.)
 Hence a definitive signature of 
 the singular structure can be obtained from the propagator by computing
 the value of the inverse square root of the propagator 1/rP$\equiv
 \frac{1}{\sqrt{P(t)}}$. Ordering the configurations according to the
 size of 1/rP should then reveal a linear extrapolation to a zero value
 at the singularity, reflecting the uniform sampling by the Monte Carlo
 in the neighborhood of the singularity. In Fig(4) we show these plots
 for the two mass values discussed above. 
\begin{figure}
\psfig{figure=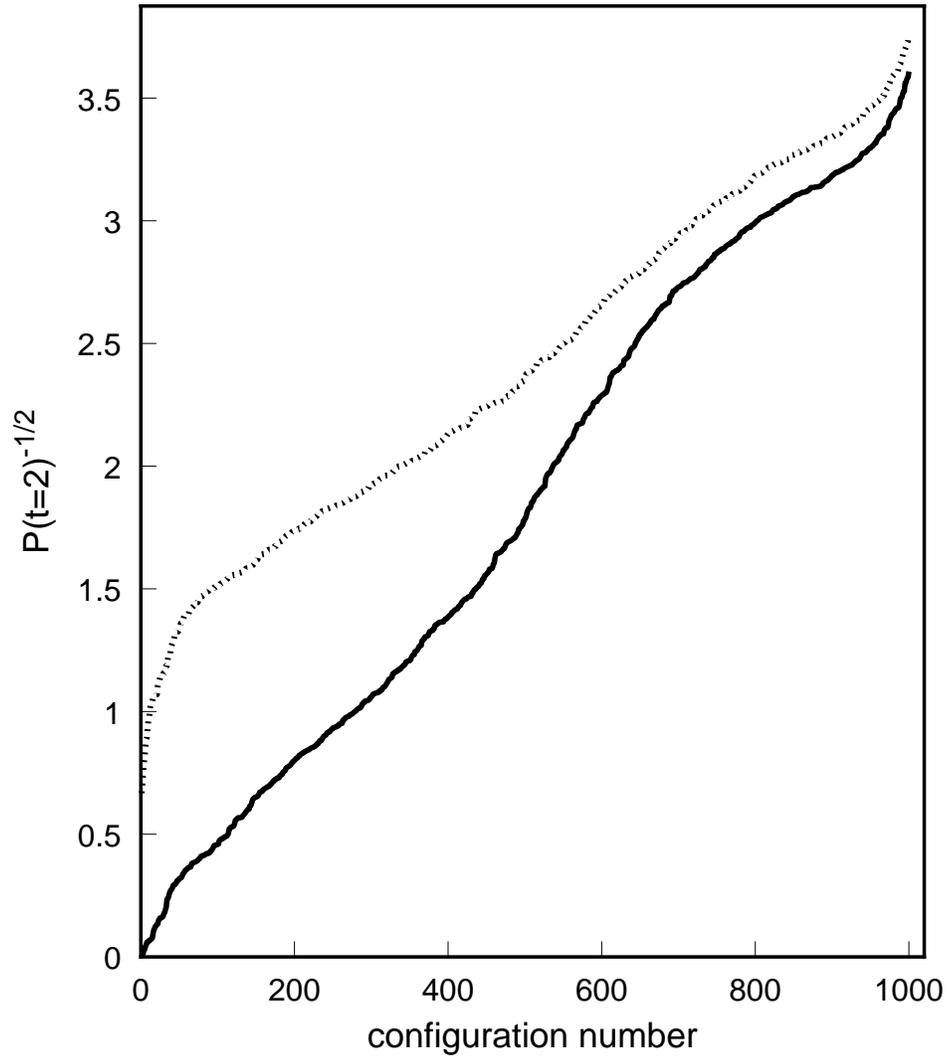,
width=0.95\hsize}
\caption{1/rP plot for 1000 configurations in QED2. The configurations are ordered
from lowest to highest value of 1/rP.  Results for $m_q = 0.10$ (dotted curve) and
$m_q = 0.06$ (solid curve) are shown.}
\label{fig:rP1000}
\end{figure}

 It is clear that the smaller
 mass value, $m_q$=0.06, shows a smooth extrapolation of 1/rP to zero
 while the larger value, $m_q$=0.10, reveals a gap.  Since the poles
 are expected to dominate only for configurations with eigenvalues close
 to the poles, only the behavior as 1/rP $\rightarrow$0 is relevant here.
 In Figure(5), we focus on this region for the small mass case to 
 bring out clearly the linear behavior close to the origin.

\begin{figure}
\psfig{figure=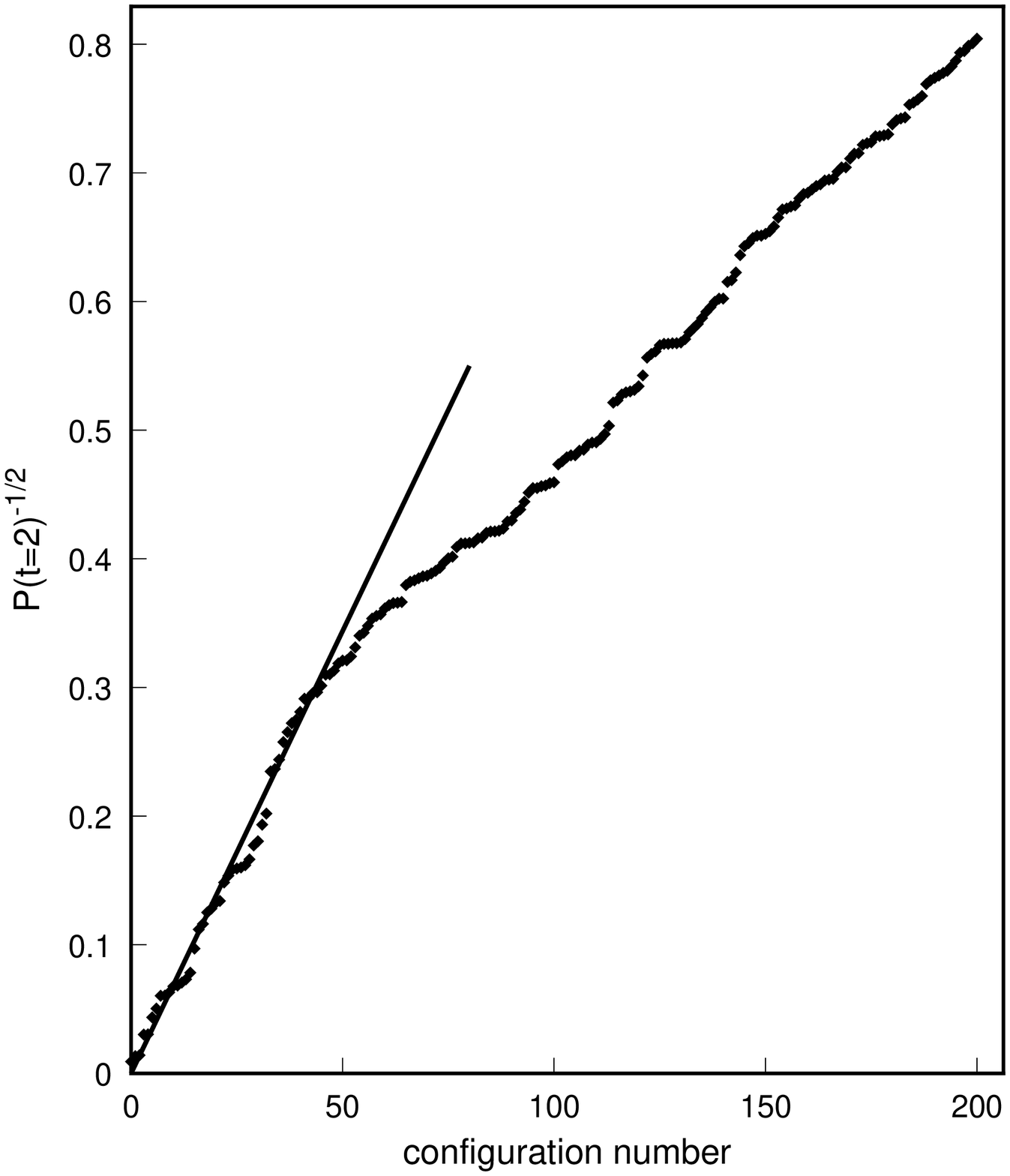,
width=0.95\hsize}
\caption{1/rP plot. Results for the lowest 200 configurations with $m_q = 0.06$ in Figure 4
are shown (small diamonds). The solid line is a linear fit to the first 40 
configurations.} 
\label{fig:rP100}
\end{figure}

\begin{figure}
\psfig{figure=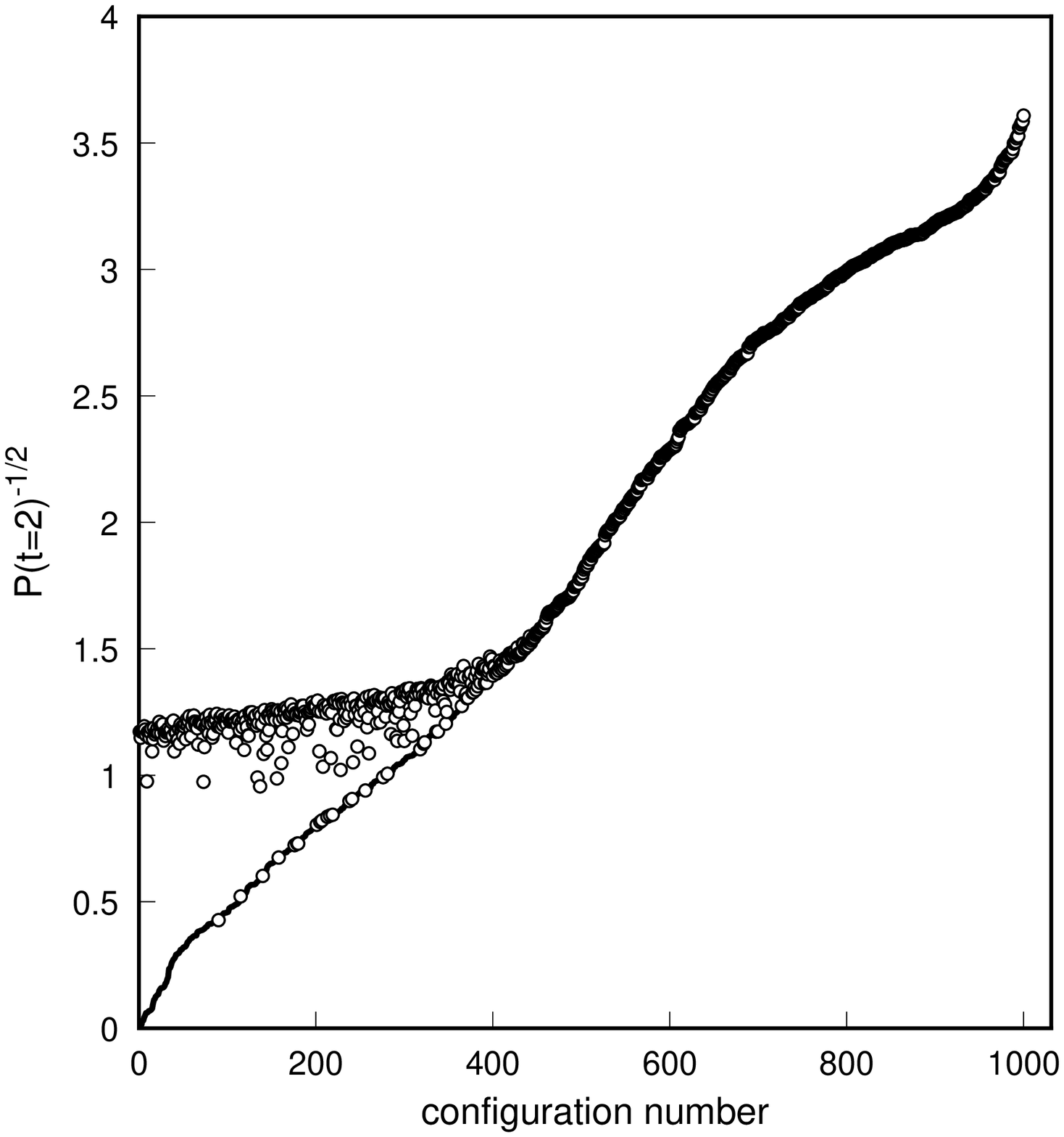,
width=0.95\hsize}
\caption{1/rP plot. The pseudoscalar correlators in Figure 4 with $m_q =0.06$ 
(solid line) are compared with the associated MQA correlators (open cirlces).}
\label{fig:rawmqa}
\end{figure}

     The correlation between the singular behavior observed in the 1/rP
plots and the real eigenvalues can be seen by comparing the raw data
with propagators modified using the MQA procedure \cite{MQA}. This
procedure modifies fermion propagators by shifting the poles due to real
eigenvalues to the critical value expected for the continuum theory.
Only real poles to the left of the critical value, as shown in Fig(1),
are shifted in this procedure. In applying the procedure to the QED2 data,
it is clear that any difference observed in the behavior of the correlation
functions is directly attributable to the structure of the real eigenvalues.
In Fig(6) we compare the 1/rP plot for the raw propagator with that for
the MQA shifted propagator. The configurations are, in each case, ordered
according to the value of the raw propagator. It is apparent that all the points
close to the origin in this plot were associated with the singular structure of the real modes.
This correlation can be made more explicit by considering 
only the configurations with real poles lying within the band shown in
Fig(1). We have checked that in QED2 these configurations all correspond to nonzero topological
charge. In Fig(7), we plot the value of 1/rP for each configuration
versus the position of the nearest real eigenvalue for that configuration. The
close correlation is obvious and the position of the singular eigenvalue
is clearly identified for this mass value, $m_q$=0.06. In many cases
(especially in QCD4)  
it is not possible to identify all the real poles. However, we have seen
that the behavior of 1/rP can clearly be used to exhibit the singularities
associated with bands of real eigenvalues. The method will only be
effective however when there is sufficient Monte Carlo statistics to
smoothly sample the neighborhood of the propagator singularities.

\begin{figure}
\psfig{figure=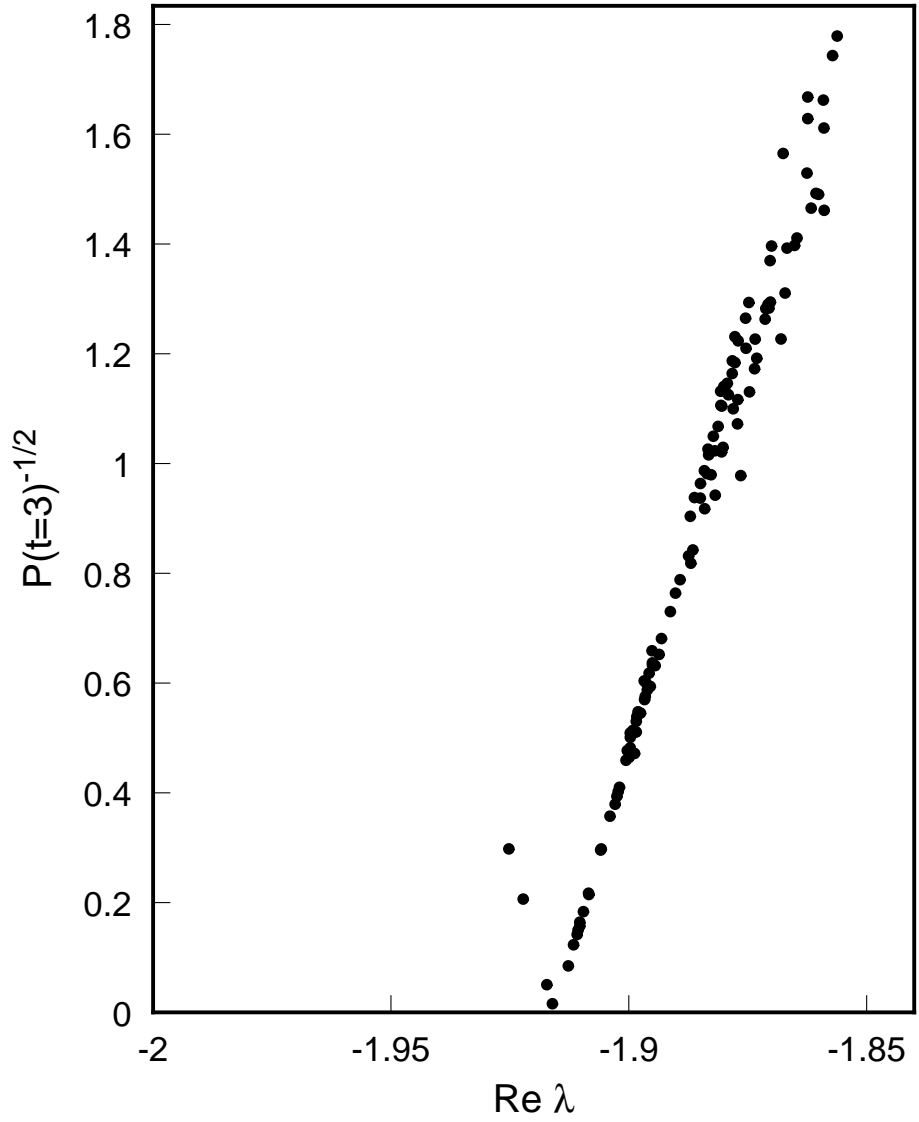,
width=0.95\hsize}
\caption{1/rP versus the position of the nearest real pole eigenvalue Re $\lambda$ 
for $m_q = 0.06$. [QED2].}
\label{fig:rP100corr}
\end{figure}

\newpage
\section{A simple test function}

    We have argued that the presence of a band of real eigenvalues of the
Wilson-Dirac operator implies that the unmodified quenched functional 
integral simply does not exist for bare fermion masses which place the
propagator singularities within the real eigenvalue band. Even with
unlimited Monte Carlo statistics, the real poles generate exceptional
configurations at a fixed rate, with fluctuations increasing as eigenvalues
appear closer to the poles of the fermion propagator. This problem can
be seen in the clearest possible way by attempting to estimate a
singular one dimensional integral, corresponding to the meson propagator,
by Monte Carlo methods. The properties of such a simulation, where we know
that the answer does not exist, can then be compared with our Monte Carlo
simulations of QED2 and QCD4.

  The test function we wish to examine is a simple integral of a double
pole singularity with a spectral weight similar to that seen for the real
eigenvalues of the QED2 analysis. We chose an integral of the form
\begin{equation}
   I = \int_{0}^{\infty} dx 2x\frac{e^{-x^2}}{(x-a)^{2}+b}
\end{equation}
For positive values of $a$ and $b$=0 the integral is divergent and we do
not expect the estimate of its value by Monte Carlo methods to converge.
Analogously to the case of QED2, we generate 1000 configurations according
to the spectral weight $2x\exp(-x^2)$. To imitate the contributions of
configurations with 
complex poles, we have randomly chosen the width parameter $b$ to be 
either zero or a fixed nonzero value for each term in the sum. As a 
result we have a sample of 200 configurations to integrate a singular
double pole term (with $b$=0) and 800 configurations with integrable
complex pole contributions (the 4:1 balance chosen here was similar to
that found in the QED2 simulation). In Fig(8) the cumulative averages and errors for
the integral are plotted as a function of the number of configurations included
(cf Fig(2)). While there seems to be an approximate convergence with 300-400
configurations, the exceptional terms associated with Monte Carlo sampling
close to the pole causes the error eventually to explode- reflecting the fact
that the integral does not exist. This plot is remarkably similar to the
plot for the pseudoscalar propagator for QED2 shown in Fig(2). In fact, the cause of 
the exceptional configurations in both cases is exactly the same - a nonintegrable dipole 
singularity- so the resemblance should hardly be surprising. We can also
plot the analogue of the 1/rP plot of Fig(4). In Fig(9) we show the inverse square
root of the individual summands placed in sequential order. The plot again
shows the linear behavior near the origin due to sampling near the pole singularity.
In addition, we have shown, in open circles, the effect of applying the
equivalent of the MQA shift moving subcritical real eigenvalues to $a$=0; again,
this removes the linear part of the curve as in the QED2 case and 
introduces a gap. We note that fluctuations in the imaginary part of
the eigenvalues and in the pole residues for QED2
explain the deviations between QED2 and
the simple test function.   However, the 
approximate quantization of the pole
residues (which are effectively instanton zero modes in the QED2 case)
is important in generating the observed
linear behavior.  From the analysis we  see that the 
1/rP plot provides a useful and clear
signal of nonintegrability in Monte Carlo simulations.

\begin{figure}
\psfig{figure=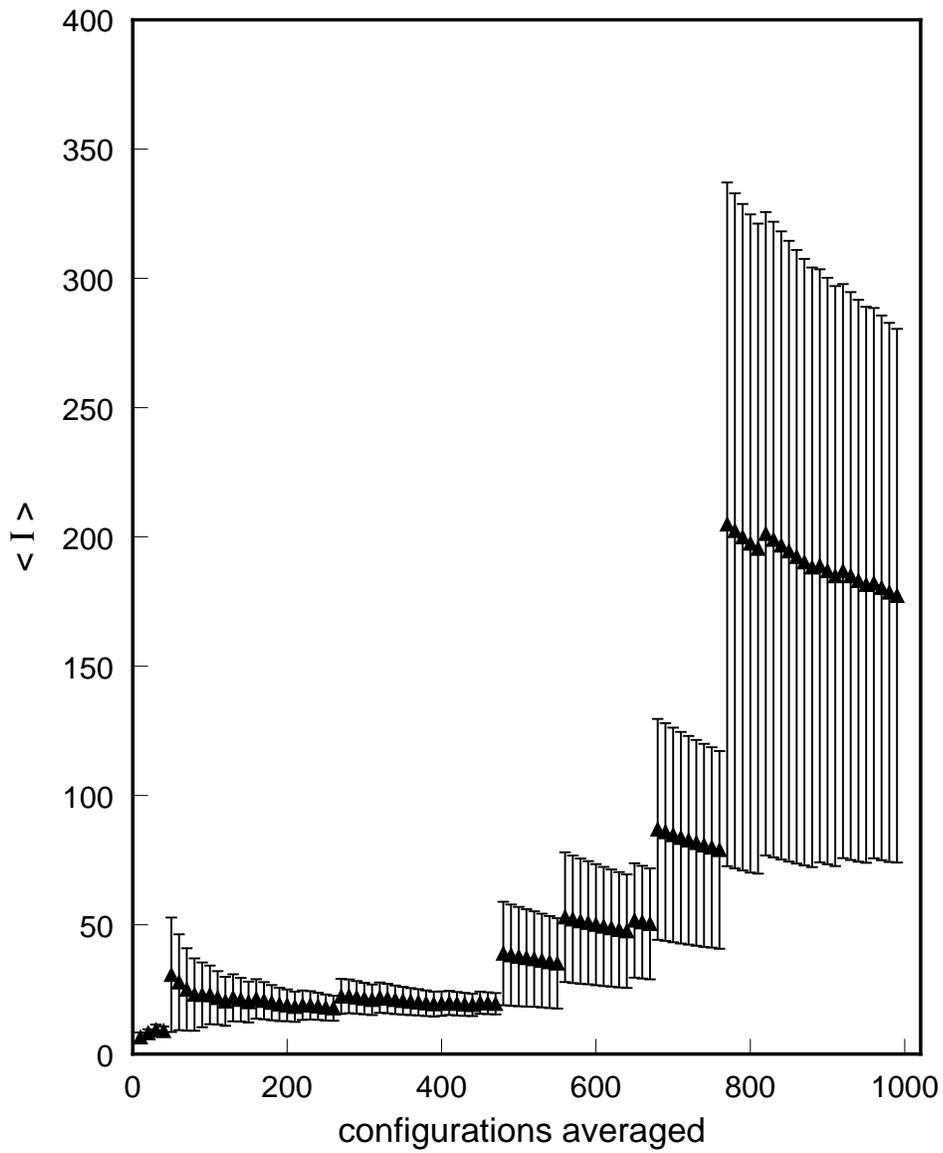,
width=0.95\hsize}
\caption{Cumulative average for Monte Carlo estimation of a test function $I$ (Eq. 4)
versus number of sampling points (``configurations'').}
\label{fig:testconv}
\end{figure}

\begin{figure}
\psfig{figure=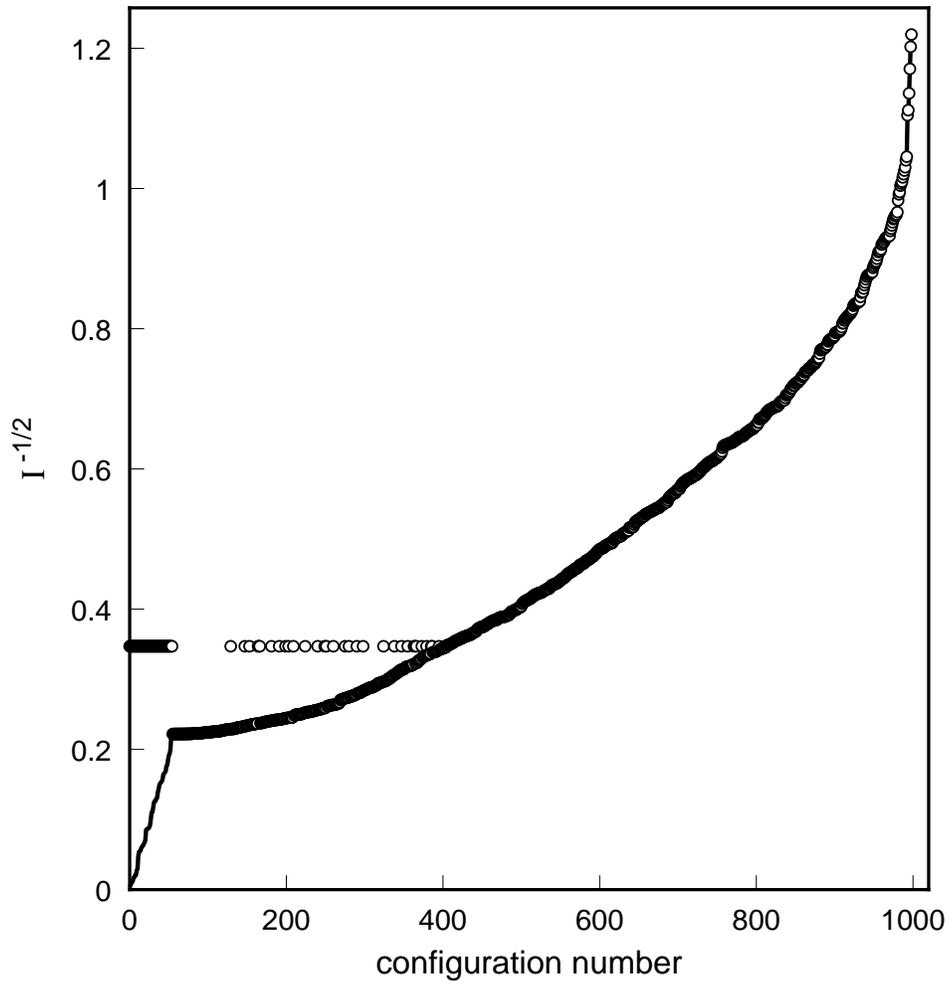,
width=0.95\hsize}
\caption{1/rP for the test function I. Solid curve represents the raw ``configurations''
ordered by value; while the open circles show the effect of applying an 
MQA procedure.}
\label{fig:rPtest}
\end{figure}

\newpage

\section{Quenched Singularity Structure of QCD4}

 Quenched calculations of quantum chromodynamics in four dimensions are also
 known to suffer from problems associated with exceptional configurations
 and singularities occasioned by real eigenvalues. In a previous paper \cite{MQA}
 we have been able to identify the spectrum of subcritical real eigenvalues
 using fits to the kappa dependence of observables such as the integrated
 pseudoscalar density. We can also study the statistical behavior of the quenched
 singularities using the 1/rP plots for the pion propagators discussed above.
 In Fig(10) we show the 1/rP plot for a 16$^3$x32 lattice at $\beta$=5.7
 and a kappa value of 0.1685, corresponding to a quark mass within the band
 of real eigenvalues for this $\beta$. There is clear evidence of the linear
 behavior near the origin. We can see that this behavior is directly 
 attributable to real eigenvalues by plotting (open circles) the same
 configurations with propagators computed using the MQA analysis where only real poles  are  
 shifted. Again, all of the contributions near the origin are removed, indicating
 that the real poles dominate the exceptional configurations. Here again,
 direct Monte Carlo simulation of the quenched functional integral will be seen
 to fail once sufficient statistics are accumulated. At heavier quark masses
 the density of real eigenvalues is smaller and a much higher level of Monte
 Carlo statistics may be needed to clearly identify the singular behavior,
 which is nevertheless still present, at least for fermion masses $m_0<$ 2.0.
\begin{figure}
\psfig{figure=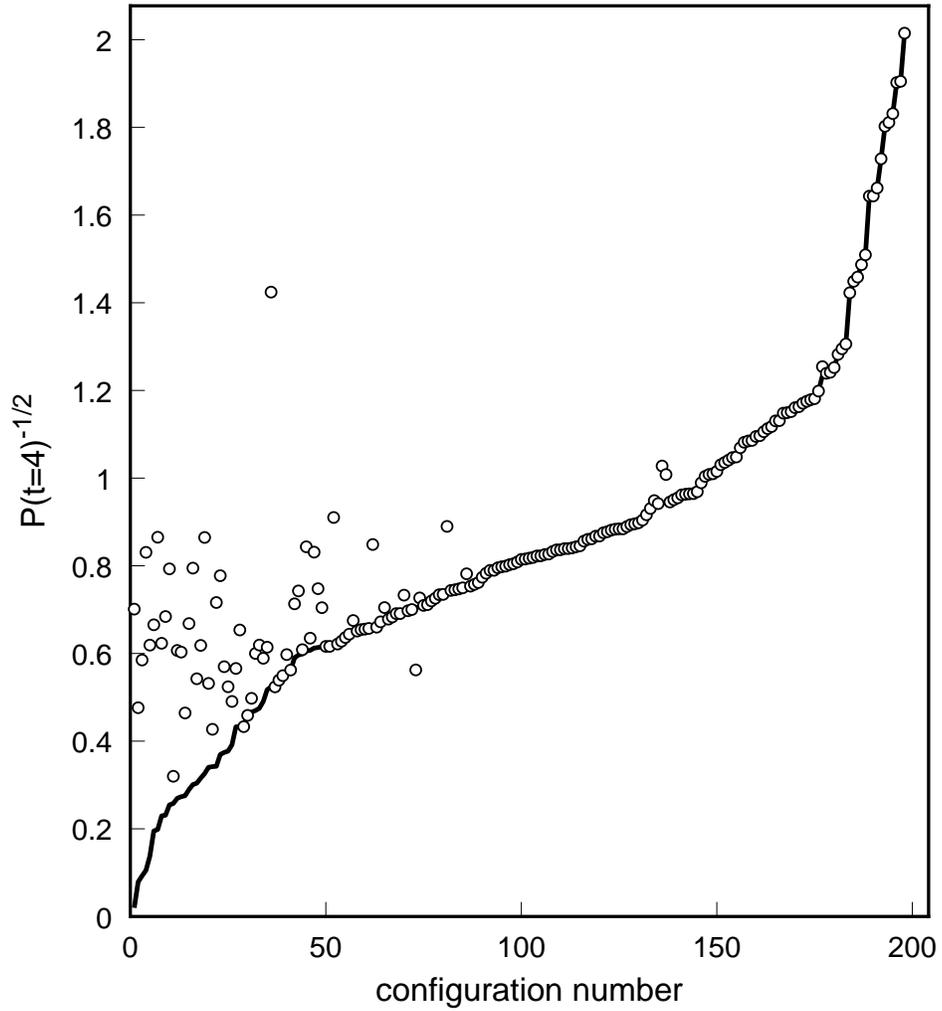,
width=0.95\hsize}
\caption{1/rP plot for 200 configurations at $\beta$=5.7 on a 16$^3$x32 lattice [QCD4].
The Wilson quark propagators used $\kappa$=0.1685.  
Configurations are ordered by value of 1/rP.
The naive pseudoscalar correlator results are denoted as a solid curve,
while the corresponding results using the MQA proceedure are denoted by open circles.}
\label{fig:rPQCD}
\end{figure}
 
  The singular behavior of the Monte Carlo integrals (deriving in turn from the
 underlying nonintegrable singularity of the quenched functional integral) is
 not a unique feature of the Wilson action. Improved actions, such as
 the Sheikholeswami-Wohlert \cite{SheikWohl} action, also suffer from the
 singularities associated with real eigenvalues. Subcritical real modes can
 be extracted from the kappa dependence of hadron propagators or we may use
 the 1/rP test to see the impact of the exceptional configurations. In Fig(11)
 we show the 1/rP plot for a 12$^3$x24 lattice at $\beta$=5.7 with clover coefficient 
 $C_{\rm sw}=$1.57 and $\kappa=$0.1425. Once more, the linear behavior near the
 origin gives a clear signal of the singular nature of the unmodified Monte Carlo
 integral and the comparison with the MQA analysis shows that the real eigenvalues
 play the essential role.

\begin{figure}
\psfig{figure=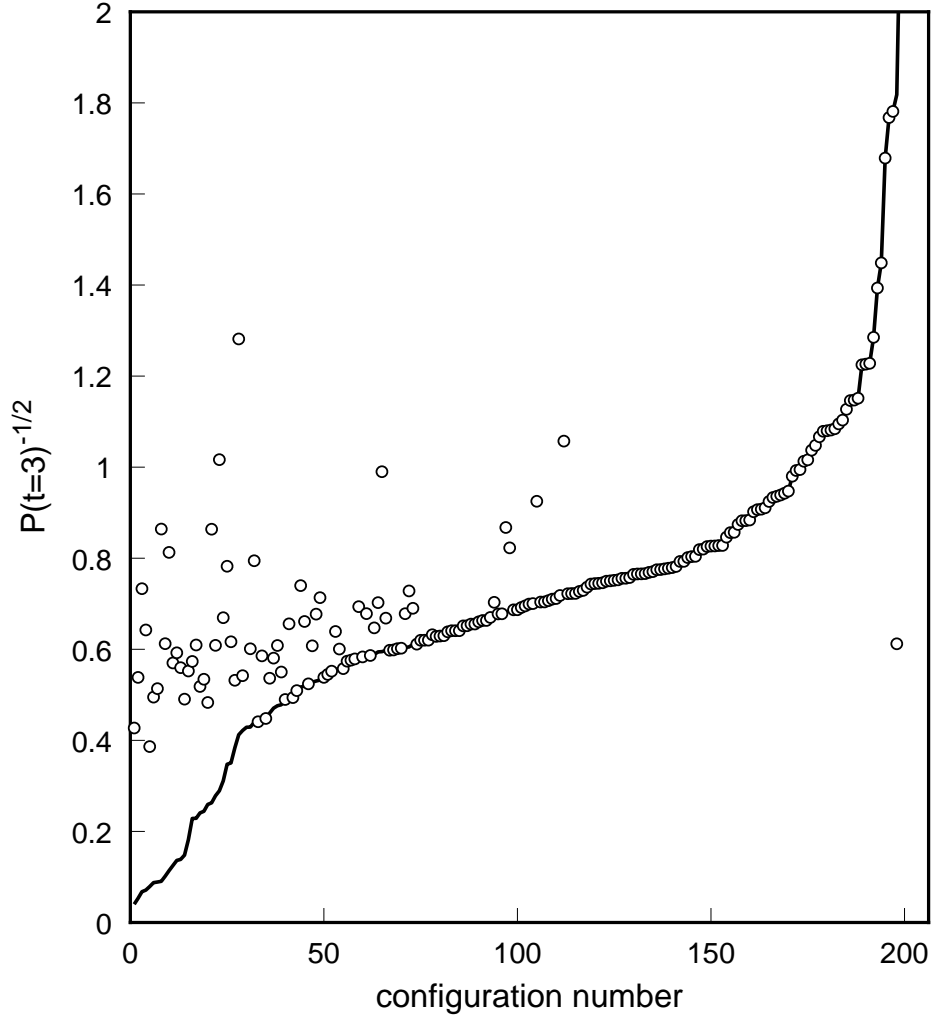,
width=0.95\hsize}
\caption{1/rP plot for 200 configurations at $\beta$=5.7 on a 16$^3$x32 lattice [QCD4].
The SW quark propagators used $\kappa$=0.1425 and $C_{\rm sw}=$1.57.
Configurations are ordered by value of 1/rP.
The naive pseudoscalar correlator results are denoted as a solid curve,
while the corresponding results using the MQA proceedure are denoted by open circles.}
\label{fig:rPQCDclover}
\end{figure}

  These results provide convincing evidence that real modes are responsible for
generating the singular behavior of the unmodified quenched approximation.
For sufficiently light quark masses, hadron propagators determined from the
valence quark propagators contain nonintegrable singularities. With low statistics
these singularities are manifested by the appearance of an occasional exceptional 
configuration. However, it is {\em inevitable} that the Monte Carlo average
will diverge as the statistics is increased and the eigenvalues near the propagator
poles are closely sampled. While we have emphasized spectral properties associated with the
Wilson-Dirac operator, it is important to realize that these
nonintegrable  singularities are
present in completely physical hadronic amplitudes in the unmodified quenched theory,
such as the pseudoscalar correlation functions.

\section{Exceptional configurations, gapless phases, and chiral source terms}

   A number of authors have preferred to use a hermitian operator $H(m)$ \cite{hermitian}
to discuss
properties of the quenched theory. This operator is simply related to the
Wilson-Dirac operator
\begin{equation}
     H(m) = \gamma_{5}(D-W)
\end{equation}
but has a completely different eigenvalue spectrum. Since $H(m)$ is hermitian it
has only real eigenvalues and (with the exception of zero eigenvalues)
it is not possible to uniquely associate 
eigenvalues of $H(m)$ with those of the Wilson-Dirac operator in a 1-1 fashion,
and therefore to identify modes which play the special role of the real 
eigenvalues in the Wilson-Dirac case. Nevertheless, the physical quantities
of the quenched theory (meson propagators and other hadronic amplitudes)
are unique and share the same singular behavior that we have established
using the Wilson-Dirac approach.

  The small eigenvalues of the $H(m)$ operator are usually identified with the
dynamical breaking of chiral symmetry. However, we  know that these
eigenvalues must also reflect the nonintegrable singularities found in the
Wilson-Dirac approach. Indeed, exactly zero eigenvalues arise from common
eigenfunctions. By varying the fermion mass we can establish the same band of
masses which correspond to singularities caused by subcritical real eigenvalues
of the Wilson-Dirac operator. Thus, the closing of a gap in the spectrum of the $H(m)$ operator corresponds in part to the appearance of nonintegrable
contributions in the quenched path integral. In Fig(12,13) we display
the central part of the spectral region of $H(m)$ in QED2 for the
bare quark value masses 0.06 and 0.10 studied earlier. The disappearance
of a gap for the smaller mass exactly corresponds to the fact that the 
rightmost cross in Fig(1) is immersed in a band of real eigenvalues of
the complex Wilson-Dirac operator.

\begin{figure}
\psfig{figure=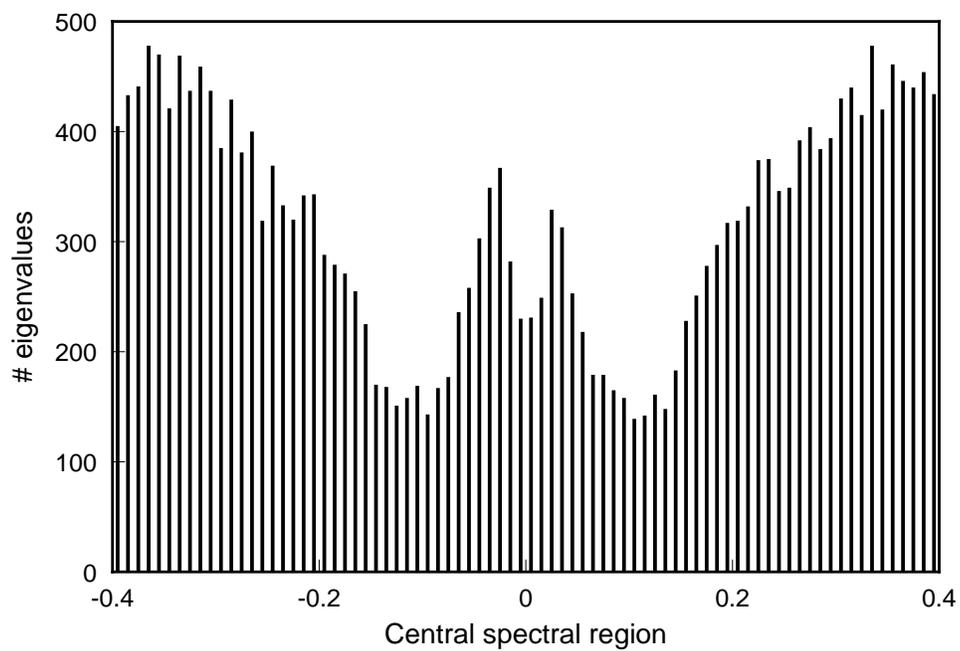,
width=0.95\hsize}
\caption{Spectral density of $H(m)$ for QED2 with $m_q=0.06$. 
The total number of eigenvalues are shown with bin size $0.01$.}
\label{fig:spec0.06}
\end{figure}

\begin{figure}
\psfig{figure=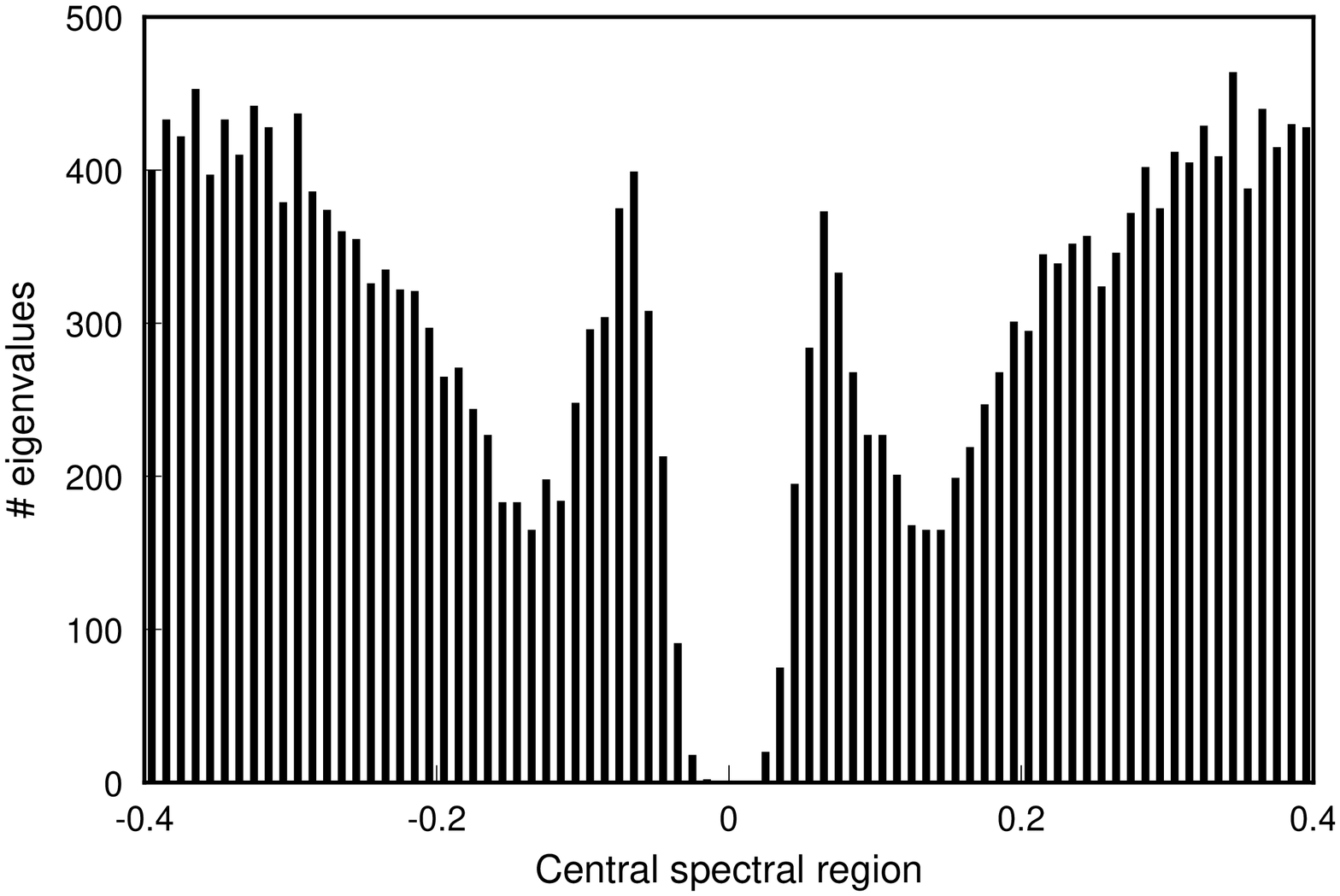,
width=0.95\hsize}
\caption{Spectral density of $H(m)$ for QED2 with $m_q=0.10$.
Bin size $0.01$.}
\label{fig:spec0.10}
\end{figure}

  Some authors \cite{Florida} have recently suggested that this band should
reflect the dynamical realization of the chiral phases of the quenched theory.
They propose an identification of the entire band with a massless phase of the
theory. These conclusions are based on studies with limited statistics on
small lattices. We do not think that their conclusions can be sustained
once larger data samples are examined. The arguments presented above show
that the normal Monte Carlo integral of the quenched theory for meson
propagators simply does not exist if the fermion masses lie within the band 
identified with  the putative massless phase . Hence, any
inference of a particular physical behavior of the meson propagators based
on a small sample of configurations, particularly exceptional configurations
where real modes are present, is unwarranted. We make no specific claims here about the
possible existence, nonexistence or properties of additional massless chiral phases 
in the unquenched theory at infinite volume. However,
for the quenched Wilson theory at finite volume, the detailed statistical analyses
presented in this paper clearly demonstrate that conventional hadron correlators (and in particular,
the pion correlators needed to establish the chiral spectrum in any putative new
phase) {\em simply do not exist} once the gapless region is entered. 

  The suggestion that the flavor-parity breaking Aoki phase 
\cite{Aoki,Sharpe} of Wilson lattice QCD be studied by perturbing the
system with a flavor-parity noninvariant source leads to an interesting point concerning
the divergence of quenched amplitudes. The addition to the theory of any nontrivial 
chiral/flavor rotation of 
the conventional mass term as a source term 
(here we are dealing with two flavors of Wilson fermions,
and $\vec{\sigma}$ are the corresponding isospin generators)
\begin{equation}
   h\bar{\psi}\psi \rightarrow h\bar{\psi}e^{i\vec{\theta}\cdot\vec{\sigma}\gamma_{5}}\psi
\end{equation}
modifies the hermitian Dirac operator by
\begin{equation}
   H(m) \rightarrow H(m)+h\gamma_{5}\cos(\theta)+ih\sin(\theta)\hat{\theta}\cdot\vec{\sigma}
\end{equation}
The additional antihermitian term (nonvanishing for any $\theta\neq 0, n\pi$) clearly renders the
spectrum of $H(m)$ complex, introducing a gap, as all eigenvalues $\lambda$ of $H(m)$
now satisfy $|\lambda|>h|\sin(\theta)|$. Consequently, pion correlators computed with 
quark propagators including the source term will now be perfectly well-defined even in
the quenched theory, as the nonintegrable poles are moved away from the integration
contour of the quenched path integral. (This corresponds to the situation in the
one-dimensional test integral of Section 3 when $b\neq 0$). 
The properties of meson correlators in the Aoki
phase can thus be studied in a well-defined way at large volume before taking the limit
$h\rightarrow 0$, as is done typically in order to isolate the condensate. Of course, at
finite volume, the singularity will return as the source term is set to zero, so as in the
case of studies of spontaneous symmetry breaking at finite volume one has to be careful to extrapolate
in an appropriate way from larger values of the source parameter. But in the absence of
such a source term in the propagator inversion, or of pole shifting as in the MQA procedure,
one will necessarily encounter completely undefined 
statistical averages of meson propagators as more configurations 
are included in a quenched simulation.

\section{Summary}
  We conclude by summarizing our basic results on the nature of the
 quenched chiral artifacts in the Wilson-Dirac formulation of lattice gauge 
 theory:\\
(1) The appearance of exceptional configurations in quenched simulations
 is directly attributable to the appearance of exactly real modes of the
 Wilson-Dirac operator in nontrivial topological charge sectors of the 
 theory. \\
(2) Even in cases where the extraction of the spectrum of the Wilson-Dirac
 operator is computationally prohibitive, there are a number of reliable 
 statistical tests that can be applied to diagnose the presence of  the exactly
 real modes. In particular, the convergence properties of
 cumulative averages of physical quantities (cf Figs (2,3,8)), as well as the
 behavior of the reordered inverse square-root of meson correlators 
 (cf Figs(4,5,6,7,9,10,11)) 
 can be used to signal directly  the appearance of such modes. \\
(3) Finally, the statistical diagnostics used here clearly reveal the
 nature of the singularity of the quenched functional integral which corresponds
 to a one-dimensional integral with a nonintegrable
 singularity. The singularity can be removed, and a well-defined quenched
 theory obtained, either by the MQA pole-shifting technique introduced in 
 our earlier papers [4,5], or by introducing a chirally rotated source at
 interim stages of the simulations.

\newpage 
{\noindent \Large \bf Acknowledgements}

  The work of W. Bardeen and E. Eichten was performed at the Fermi National
Accelerator Laboratory, which is operated by University Research Association,
Inc., under contract DE-AC02-76CHO3000. A. Duncan is grateful for the
hospitality of the Fermilab Theory Group, where this work was performed.
The work of A. Duncan was supported in part by NSF grant PHY97-22097.
The work of H. Thacker was supported in part by the Department of Energy
under grant DE-AS05-89ER40518.

\end{document}